\documentclass[12pt, prd, showpacs, nofootinbib]{revtex4}

\usepackage{amsfonts,amssymb,amsmath}

\newcommand{\bls}[1]{\renewcommand{\baselinestretch}{#1}}
\def\noi{\noindent}

\def\nhq{\hspace*{-0.5em}}

\def\cm{\hspace*{1cm}}

\def\wide{\mbox{$\dst\vphantom{\int}$}}

\def\Acknow#1{\subsection*{Acknowledgment} #1}

\def\Jl#1#2{#1 {\bf #2},\ }

\def\ApJ#1 {\Jl{Astroph. J.}{#1}}
\def\CQG#1 {\Jl{Class. Quantum Grav.}{#1}}
\def\DAN#1 {\Jl{Dokl. AN SSSR}{#1}}
\def\GC#1 {\Jl{Grav. Cosmol.}{#1}}
\def\GRG#1 {\Jl{Gen. Rel. Grav.}{#1}}
\def\JETF#1 {\Jl{Zh. Eksp. Teor. Fiz.}{#1}}
\def\JETP#1 {\Jl{Sov. Phys. JETP}{#1}}
\def\JHEP#1 {\Jl{JHEP}{#1}}
\def\JMP#1 {\Jl{J. Math. Phys.}{#1}}
\def\NPB#1 {\Jl{Nucl. Phys. B}{#1}}
\def\NP#1 {\Jl{Nucl. Phys.}{#1}}
\def\PLA#1 {\Jl{Phys. Lett. A}{#1}}
\def\PLB#1 {\Jl{Phys. Lett. B}{#1}}
\def\PRD#1 {\Jl{Phys. Rev. D}{#1}}
\def\PRL#1 {\Jl{Phys. Rev. Lett.}{#1}}

\def\al{&}
\def\lal{&&\nhq {}}
\def\eq{Eq.\,}
\def\eqs{Eqs.\,}
\def\beq{\begin{equation}}
\def\eeq{\end{equation}}
\def\bear{\begin{eqnarray}}
\def\bearr{\begin{eqnarray} \lal}
\def\ear{\end{eqnarray}}
\def\earn{\nonumber \end{eqnarray}}
\def\nn{\nonumber\\ {}}

\def\yy{\\[5pt] {}}
\def\yyy{\\[5pt] \lal }
\def\eql{\al =\al}

\def\dst{\displaystyle}

\def\d{\partial}

\def\sign{\mathop{\rm sign}\nolimits}
\def\diag{\mathop{\rm diag}\nolimits}

\def\const{{\rm const}}

\def\then{\ \Rightarrow\ }

\def\N{{\mathbb N}}

\def\mn{_{\mu\nu}}
\def\mN{_{\mu}^{\nu}}
\def\MN{^{\mu\nu}}

\def\vac{{}^{\rm vac}}
\def\ebh{e_{\rm bh}}

\def\sph{spherically symmetric}
\def\ssph{static, spherically symmetric}
\def\bh{black hole}
\def\bhs{black holes}
\def\RN{Reissner--Nordsrtr\"om}

\bls{1.08}
\begin{document}
%%\onecol

\title{Neutral and charged matter in equilibrium with black holes}

\author{K. A. Bronnikov}
\affiliation{Center for Gravitation and Fundamental Metrology, VNIIMS, 46 Ozyornaya
Street, Moscow 119361, Russia, and\\ Institute of Gravitation and Cosmology,
PFUR, 6 Miklukho-Maklaya Street, Moscow 117198, Russia}
\email{kb20@iyandex.ru}

\author{O. B. Zaslavskii}
\affiliation{Department of Physics and Technology, Kharkov V.N. Karazin National
University, 4 Svoboda Square, Kharkov, 61077, Ukraine}
\email{zaslav@ukr.net}

\begin{abstract}
  We study the conditions of a possible static equilibrium between \sph,
  electrically charged or neutral \bhs\ and ambient matter. The following
  kinds of matter are considered: (1) neutral and charged matter with a
  linear equation of state $p_r = w\rho$ (for neutral matter the results of
  our previous work are reproduced), (2) neutral and charged matter with
  $p_r \sim \rho^m$, $m > 1$, and (3) the possible presence of a ``vacuum
  fluid'' (the cosmological constant or, more generally, anything that
  satisfies the equality $T^0_0 = T^1_1$ at least at the horizon). We find a
  number of new cases of such an equilibrium, including those generalizing
  the well-known Majumdar-Papapetrou conditions for charged dust. It turns
  out, in particular, that ultraextremal \bhs\ cannot be in equilibrium with
  any matter in the absence of a vacuum fluid; meanwhile, matter with $w >
  0$, if it is properly charged, can surround an extremal charged \bh.  
\end{abstract}

\maketitle

%%%%%%%%%%%%%%%%%%%%%%%%%%%%%%%%%%%%%%%%%%%%%%%%%%%%%%%%%%

\section{Introduction}

  Black holes under real astrophysical conditions are always surrounded by 
  some kind of matter which is either in equilibrium with the \bh\ or is 
  falling on it. It is therefore of great interest to know the conditions 
  of such an equilibrium and, in particular, which kinds of matter are 
  compatible with it.

  In our previous papers \cite{curly, nonsph} we studied the static 
  equilibrium conditions between a black hole and ambient matter with the
  linear equation of state $p_r  = w\rho$, where $\rho$ is the density,
  $w=\const$, and $p_r$ is the pressure in the direction orthogonal to the
  horizon (radial pressure in the case of spherical symmetry). It is well
  known that such an equilibrium is possible in the case $w = -1$, and
  examples of the corresponding space-times are those with a cosmological
  constant and electromagnetic fields (e.g., the \RN--de Sitter space-time).
  According to \cite{curly, nonsph}, an equilibrium is also possible for
  some other values of $w$, forming a discrete set ($w = -1/3$ in the
  generic case, which corresponds to disordered cosmic strings if we
  additionally suppose that matter is isotropic).

  In \cite{nl}, the same problem was considered for nonlinear equations of
  state of the type $p_r +\rho  = w_h (\rho -\rho_h)^{\beta}$ where $\beta
  >1$ and $\rho_h$ is a nonzero density at the horizon.
  It was shown there that the results for isotropic and anisotropic fluids
  are very different. In the first case, the conditions for matter to be in
  equilibrium with the horizon are direct generalizations of those previously
  found \cite{curly}. In the second case, it turned out that there is no
  restriction on the equation of state but the horizon type is severely
  restricted: it should be simple only.

  In this paper we reconsider this problem including matter with another type 
  of nonlinear equations of state, $p_r \sim \rho^m$, $m > 1$, which is of
  particular astrophysical interest since it includes polytropic equations
  of state. We also include into consideration electrically charged black
  holes and the possible electric charge of matter itself.

  Although the electric charges of celestial bodies are usually negligible
  (see, e.g., \cite{baush09}), this is probably true not for all objects;
  in particular, there are indications that neutron and quark stars can
  have quite significant charge to mass ratios and huge charges up to 
  $10^{20}$ C \cite{jose04}, and there are attempts to build models 
  of astrophysical objects where the electric charge plays a substantial 
  role \cite{e-lin, e-nlin}. Moreover, a significant electric charge is 
  important for the formation of so-called quasiblack holes, systems which 
  are on the threshold of forming a horizon (see \cite{LZ-07} and references 
  therein). 

  On the other hand, qualitative effects are possible which depend on the 
  very existence of such a charge and taking place even at its very small 
  values. In \bh\ physics, electric (or magnetic) charges can lead to 
  results of general theoretical interest --- such as the famous 
  Majumdar-Papapetrou solutions \cite{maj-pap} describing an equilibrium 
  between charged extremal black holes and dustlike matter with the charge 
  density $\rho_e = \pm\rho$ in proper units. Let us note that such an 
  equilibrium (also important for quasiblack hole models) can be achieved 
  even if a sphere of neutral hydrogen loses a fraction of $10^{-18}$ of 
  its electrons.

  It is also known that many features are qualitatively similar in rotating
  and electrically charged black holes. The first case is of greater
  interest in astrophysics but is more complicated. Therefore, it looks
  reasonable to study \sph\ configurations of Einstein-Maxwell systems to
  gain experience for future studies of rotating black holes.

  The paper is organized as follows. In Sec.\,II we present the basic
  equations and conditions to be satisfied on a horizon. In Sec.\,III, we
  consider the following opportunities for matter near both charged and
  neutral spherically symmetric black holes: (1) neutral and charged matter
  with a linear equation of state $p_r = w\rho$ (for neutral matter the
  results of \cite{curly} are reproduced), (2) neutral and charged matter
  with $p_r \sim \rho^m$, $m > 1$, and (3) the existence of a possible
  ``vacuum fluid'' (the cosmological constant or, more generally, anything
  that satisfies the condition $p_r\vac + \rho\vac =0$ \cite{dym}, at least
  at the horizon) and its influence on the equilibrium conditions. We will
  find a number of new cases of such an equilibrium, in particular, those
  generalizing the Majumdar-Papapetrou conditions. Section IV is a conclusion.

\section{Basic equations}

\subsection{Field equations}

  Consider a generic spherically symmetric space-time with the metric
  	(We use the units $c= \hbar = G =1$)
\beq
	ds^2  = A(u) dt^2 - \frac{du^2}{A(u)} - r^2(u) d\Omega^2 .
\eeq
  Here, $u$ is the so-called quasiglobal coordinate especially suitable for
  dealing with Killing horizons because at such horizons it behaves as
  Kruskal-like coordinates in which the metric is manifestly regular there
  (see \cite{book, we09} and references therein). For this reason, we can
  require analyticity of the metric at horizons in terms of $u$ (so that $A
  \sim u^n$, $n\in \N$ if $u=0$ is a horizon) remaining in the framework of
  a static metric.

  We consider matter at rest, with the stress-energy tensor (SET)
\beq
       T\mN{}^m = \diag (\rho,\ -p_r,\ -p_\bot, -p_\bot),     \label{T_m}
\eeq
  and a radial electric field with the SET
\beq                                                          \label{T_em}
       T\mN{}^{\rm em} = \frac{e^2}{8\pi r^4} \diag (1,\ 1,\ -1,\ -1),
\eeq
  where $e = e(u)$ is the charge inside a sphere of radius $r = r(u)$. The
  matter is supposed, in general, to be electrically charged, so the tensors
  (\ref{T_m}) and (\ref{T_em}) are not conserved each separately, but only
  their sum is.

  The Maxwell equations read
\beq
	\frac{1}{\sqrt{-g}} \d_\mu (F\MN \sqrt{-g}) = 4\pi j^{\mu},
		\cm
	j^\mu = \rho_e u^\mu,                              \label{Max}
\eeq
  where $j^\mu$ is the current, $\rho_{e}$ is the invariant charge density,
  and $u^\mu$ the 4-velocity of matter. For matter at rest,
\beq
    	j^{\mu} =  A^{-1/2} \rho_e (1,\ 0,\ 0,\ 0).
\eeq
  Then, the only nontrivial component of the Maxwell equations has the form
\beq
      (F^{01}r^2)' = 4\pi r^2 \rho_e/\sqrt{A},   	\label{Max0}
\eeq
  where the prime stands for $d/du$, and its solution is
\beq
        F^{01} = e(u)/r^2,	  			\label{F01}
\eeq
  with
\beq
	e(u) = \ebh + e_m (u),  \cm
	e_{m}(u) = 4\pi \int_0^{u}\frac{du}{\sqrt{A}}r^2 \rho_e, \label{em}
\eeq
  where we assume (without loss of generality) that $u = 0$ is the event
  horizon of a black hole, $u > 0$ outside it and $\ebh$ is the \bh\ charge.
  The quantity $e_m (u)$ is the total charge of matter between the horizon
  and a sphere at given $u > 0$.

  The conservation law can be written as follows:
\beq
	p'_r + \frac{A'}{2A}(p_r +\rho) + \frac{2r'}r (p_r - p_\bot)
		-\frac{e\rho_e}{r^2 \sqrt{A}} = 0.    		\label{cons}
\eeq

  The Einstein equations give
\bear
	G_1^1 -G_0^0 \eql -2A\frac{r''}r  = 8\pi (p_r + \rho),   \label{01}
\yy
	G_1^1 \eql \frac 1 {r^2}[-1 + A'rr' + A r'{}^2]
				= 8\pi p_r - \frac{e^2}{r^4},   \label{11}
\yy                                                             \label{02}
	G^0_0 -G^2_2 \eql \frac{1}{2r^2}[-2 +A (r^2)'' - A''r^2]
		  	= - 8\pi (\rho+ p_\bot) - \frac{2e^2}{r^4}
\ear

\subsection{Conditions near the horizon}

  As already mentioned, we place a horizon at $u=0$; then, since it is by
  definition a regular surface, we require analyticity of the metric, and
  hence of the SET components, in terms of the quasiglobal coordinate $u$ 
  at $u=0$. It then follows that, as $u\to 0$,
\bearr
	A\approx A_0 u^n,  \cm n > 0, 		 		\label{A}
\yyy
	\rho \approx \rho_0 u^k, \cm k \geq 0,       	        \label{rho}
\ear
  with integers $n$ and $k$. The pressure components must also
  behave as (\ref{rho}) with integer powers maybe other than $k$.

  \eq (\ref{01}) leads to the well-known result that a nonzero density of
  matter at the horizon (that is, $k=0$) is only possible if $\rho + p_r =0$
  or equivalently $T^0_0 = T^1_1$, which characterizes the so-called
  ``vacuum fluid'' \cite{dym}. If this relation holds in the whole space or
  at least in a finite region, \eq (\ref{01}) takes the form $r''=0$, and
  (provided $r\ne \const$) after evident rescaling the quantity $r$ may be
  appointed the radial coordinate, $u = r$. The metric function $A(r)$ then
  remains arbitrary and can contain zeros (horizons) of any order;
  substitution of $A(r)$ into the Einstein equations and the conservation
  law yield expressions for $\rho(r)$ and $p_\bot(r)$. Moreover, if there
  is a radial electric field, it is natural to assume that the vacuum fluid
  itself is not charged, $\rho_e = 0$, hence $e = \ebh = \const$: the \bh\
  can be charged, which does not prevent the procedure of obtaining a
  solution with given $A(r)$ because the electric field also satisfies the
  condition $T^0_0 = T^1_1$.

  Thus the situation with a vacuum fluid does not require a further study.
  It can happen, however, that $\rho + p_r = 0$ only at the horizon, as is
  the case, e.g., in \bh\ solutions with scalar fields \cite{bh-sc}\footnote
	{For a minimally coupled scalar field $\phi$, the kinetic part of
	 the SET $T\mN$ is proportional to $A\phi'{}^2$ and vanishes at
	 horizons due to $A=0$, whereas the potential enters into the SET 
	 like the cosmological constant, i.e., is vacuum-like.}
  and horizons with such matter (to be called ``pseudo-vacuum'') deserve
  separate attention. Not being vacuum in general, such matter can be both
  neutral and charged.

  Let us also adopt the following assumptions:
\begin{description}
\item[(i)]
    $|p_r/\rho| < C_1=\const <\infty$, $|p_\bot/\rho| < C_2 =\const< \infty$,
    which is a weakened version of the dominant energy condition.
\item[(ii)]
    $|\rho_e/\rho| < C_3 = \const < \infty$, and one can assume
    that near the horizon
  \beq                                                     \label{rho_e}
       \rho_e \approx \alpha \rho u^s \sim u^{k + s}, \ \ \
       		\alpha, s =\const, \ \ s \geq 0.
  \eeq
\end{description}

  To sum up, we require regularity of the metric, so that it can be
  extended through the horizon, and, using the field equations, try 
  to find out which kinds of matter are compatible with this regularity. 
  We also assume that this matter is ``not too exotic'', that is, (i) 
  the pressure to density ratio and (ii) the charge to mass density ratio 
  are bounded above.

  In what follows, we will distinguish (pseudo-)vacuum and ordinary
  matter. For ordinary matter, since $\rho + p_r \ne 0$ in general but 
  $\rho + p_r =0$ at the horizon, we must require that both $\rho$ and 
  $p_r$ vanish at the horizon, and in \eq (\ref{rho}), $k > 0$.

  We will consider equilibrium conditions between a horizon and ambient
  matter in the following cases: (1) ordinary matter with $p_r \ne -\rho$
  (neutral or charged, with linear and nonlinear equations of state), for
  which $k >0$ in (\ref{rho}), (2) pseudo-vacuum matter, for which $k=0$,
  and (3) a non-interacting mixture of ordinary and (pseudo-)vacuum matter.

%%%%%%%%%%%%%%%%%%%%%%%%%%%%%%%%%%%%%%%%%%%%%%%%%%%%%%%%%%%%%%%%%%%%%%%
\section{Matter in equilibrium with the horizon}

%%%%%%%%%%%%%%%%%%%%%%%%%%%%%%%%%%%%%%%%%%%%%%%%%%
\subsection {Charged and neutral matter with a linear equation of state}

  Suppose a linear equation of state for matter,
\beq                                                     \label{w}
	p_r = w\rho, \cm w = \const \ne -1.
\eeq
  After substitution to (\ref{cons}) we have
\beq                                                     \label{cons1}
     2A w \rho' + A' (w+1) \rho + 2\frac{r'}{r}(p_r - p_\bot)
     		- \frac{\sqrt{A}e \rho_e}{r^2} =0.
\eeq

  From (\ref{01}) it now follows $\rho(0) =0$, so that $k>0$ and, even more
  than that, $k \geq n$ (generically $k=n$, while $k > n$ corresponds to
  $r''(0) =0$).

  It is also clear from (\ref{rho_e}) and (\ref{em}) that at small $u$
\beq
      e_m \sim u^{k-n/2+s+1},                           \label{ord-em}
\eeq
  and the exponent here is manifestly positive, so that, as should be the
  case in a regular configuration, $e_m(0) = 0$.

  Let us consider separately the cases of neutral ($\ebh=0$) and
  charged ($\ebh \ne 0$) \bhs.

\medskip\noi
{\bf (a)} $\ebh  = 0$.
  It follows from (\ref{11}) that $n = 1$, since, in the order $O(1)$, the
  minus unity in the square brackets can be compensated only by the term
  $A'rr'$ and only in the case $n=1$, see \cite{curly}.

  In (\ref{cons1}) the first two terms have the order $u^{n+k-1}$, the third
  one $\sim u^{n+k}$ and the fourth one
\beq
	  e_m \rho_e \sqrt{A} \sim u^{2k + 2s + 1}.
\eeq
  Since $k\geq n$ and $s\geq 0$, the last two terms are negligible as
  compared to the first two, and we return to the old result for
  uncharged matter in the absence of a vacuum fluid \cite{curly}:
\beq
	w = -\frac 1 {1+2k}.  			        \label{w1a}
\eeq
  In the generic case $k = n = 1$ we obtain $w = -1/3$, which corresponds to
  a gas of disordered cosmic strings.

  Thus near a neutral \bh\ a possible charge of matter itself has no effect.

\medskip\noi
{\bf (b)} $\ebh \neq 0$.
  Then, (\ref{cons}) gives
\beq                                                     \label{cons1b}
     [2 kw + n (w+1)] u^{k - 1} + O (u^{k})
     	- \frac{2\alpha \ebh}{r_h^2 \sqrt{A_0 }}u^{k - n/2 + s} = 0.
\eeq

  In the case of a simple horizon ($n = 1$), the last term in (\ref{cons1b})
  is again negligible, and we return to the old condition (\ref{w1a}). In
  addition, \eq (\ref{11}) in the order $O(1)$ leads to the relation
\beq
	-1 + r r' A'\Big|_{u=0} + \ebh^2/r_h^2 =0.       \label{c1b}
\eeq
  Moreover, to provide a correct Taylor expansion in (\ref{cons1b}),
  we should require that $s$ should be half-integer.

  In the case of a double horizon, $n=2$, \eqs (\ref{11}) and (\ref{02})
  lead to
\beq                                                    \label{1b,rh}
  	r_h = |\ebh|, \cm  A_0 = 1/r_h^2.
\eeq
  Substituting all this into (\ref{cons1b}) under the assumption $s=0$
  (that is, $\rho_e/\rho = \alpha$), we obtain
\beq                                                     \label{w1b}
	w = \frac{\alpha' -1}{k+1},
\eeq
  where $\alpha' = \alpha \sign \ebh$. In the generic case $k = n = 2$, this
  gives $w = (\alpha' -1)/3$.  In particular, if $\alpha' = 1$, then $w =0$,
  and we recover the well-known Majumdar-Papapetrou configuration of
  an extremal black hole and charged dust with $|\rho_e| =\rho$.

  If matter and the black hole have charges of opposite signs, so that
  $\alpha' < 0$, then by (\ref{w1b}) we must have $w < -1/(k+1)$.

  There is one more opportunity with $n=2$: if $s >0$, then the last term
  in (\ref{cons1b}) is negligible, and we obtain the same condition as
  for neutral matter in \cite{curly} for $n=2$, namely, $w = -1/(1+k)$.
  In other words, a small charge density such that $\rho_e/\rho \to 0$
  as $u\to 0$ is insignificant for the equilibrium conditions.

  If we assume $n >2$, then \eq (\ref{11}) again gives $r_h^2 = \ebh^2$,
  but now in \eq (\ref{02}), in the order $O(1)$, we have $-1/r_h^2$ on the
  left-hand side and $-2/r_h^2$ on the right-hand side. Therefore the case
  $n > 2$ is ruled out.

%%%%%%%%%%%%%%%%%%%%%%%%%%%%%%%%%%%%%%%%%%%
\subsection{Neutral matter with a nonlinear equation of state}

  Let us assume that $\rho_e = 0$, hence $e_m =0$, but admit $\ebh \ne 0$
  (a charged black hole) and take the equation of state in the form
\beq
      p_r = B \rho^m,                                      \label{m}
\eeq
  where $m >1$, since $m <1$ would violate assumption (i) while $m=1$
  has already been considered. We discard, in particular, all equations
  of state like that of Chaplygin gas ($m < 0$ since, as $\rho\to 0$, they
  give an infinite pressure incompatible with regularity of the horizon.
  It follows from (\ref{rho}) that $p_r \sim u^{mk}$, hence $mk$
  should be a positive integer.

  The form of the conservation law (\ref{cons}) now does not depend on
  whether the black hole is charged or not. Substituting in (\ref{cons})
  $\rho\sim u^k$ and neglecting all terms of orders $o(u^{k-1})$,
  we obtain
\beq       	                                              \label{B2}
	B m\rho^{m-1}\rho' + \frac{n}{2u}\rho = 0,
\eeq
  and it follows from comparison of the exponents that
\beq
       (m-1) k + k-1 = k-1 \ \then \ (m-1)k =0,
\eeq
  which is incompatible with our assumptions.

  We see that neutral matter with the nonlinear equation of state (\ref{m}),
  irrespective of the presence or absence of an electromagnetic field,
  cannot be in equilibrium with a black hole.

%%%%%%%%%%%%%%%%%%%%%%%%%%%%%%%%%%%%%%%%%%%
\subsection{Charged matter with a nonlinear equation of state}

  Let us discuss the behavior of electrically charged matter with the
  equation of state (\ref{m}). Neglecting the manifestly negligible terms,
  we can write the conservation law near the horizon as follows:
\beq                                                        \label{cons3}
      B m \rho^{m-1}\rho' + \frac{A'}{2A} \rho
     				- \frac{e \rho_e}{r^2\sqrt{A}} =0.
\eeq

  We again consider separately the cases of charged and uncharged black
  holes.

\medskip\noi
{\bf (a)} $\ebh = 0$.
  As in Sec. IIIA, it follows from (\ref{11}) that $n = 1$.

  In (\ref{cons3}), the orders of magnitude in the three terms are
\beq
	u^{km-1}, \ \ \ u^{k-1}, \ \ \  u^{2k + 2s},     \label{ord3a}
\eeq
  respectively, so that the second term, $O(u^{k-1})$, is the largest and
  cannot be compensated by other terms. Thus charged nonlinear matter cannot
  be in equilibrium with a neutral black hole.

\medskip\noi
{\bf (b)} $\ebh \ne 0$.
  Now \eq (\ref{11}) does not restrict the order of the horizon.
  In (\ref{cons3}), the first two terms are of the same orders of magnitude
  as in (\ref{ord3a}), but the third one is now $\sim u^{k+s-n/2}$.
  Therefore an equilibrium is possible if
\beq                                                        \label{ord3b}
	n/2 - s = 1.
\eeq
  A simple horizon ($n=1$) is impossible since $s\geq 0$, a double horizon
  requires $s=0$ and so on.

  For $n = 2$, just as in Sec. III.A, \eqs (\ref{11}) and (\ref{02}) lead to
  $\ebh^2 = r_h^2$ and $A_0 = 1/r_h^2$. Then from (\ref{cons3}) we find that
\beq
	\alpha = \sign \ebh = \pm 1.                         \label{al-c}
\eeq
  This means that a double horizon of a charged \bh\ can be in equilibrium
  with any matter obeying (\ref{m}) ($m > 1$) whose charge density
  $\rho_e = \pm \rho$ has the same sign as the \bh. The only restriction on
  $m$ follows from the analyticity requirement for $p_r$ (see Sec. II.B)
  according to which $m k \in \N$.

  Horizons of orders $n > 2$ are impossible precisely for the same reason as
  in Sec. III.A.

\subsection{Pseudo-vacuum matter}

  Suppose that near the horizon all relevant functions are represented by
  Taylor series whose first terms are
\bear                                                      \label{ps1}
       \rho \eql \rho_h + c_1 u^a + \ldots, \ \ a = 1,\ 2, \ldots;
\nn
      	p_r \eql -\rho_h + w_1(\rho -\rho_h)^q + \ldots,\ \ q = 1,\ 2,\ldots;
\nn
     	p_\bot \eql p_r + c_2 u^b + \ldots, \ \ b = 0,\ 1,\ 2, \ldots,
\ear
  where $a$ and $q$ are generically equal to unity and $c_1,\ c_2,\ w_1$ are
  constants. A true vacuum fluid, in which $p_r \equiv -\rho$, corresponds
  to the special case $w_1 =-1$, $q=1$. Moreover, a pseudo-vacuum matter can
  be charged, and then instead of \eq (\ref{rho_e}) we have
\beq                                                    \label{rho_e1}
	\rho_e \approx \alpha \rho_h u^s  \ \ \then \ \  e_m \sim u^{s+1},
\eeq
  where $\alpha,\,s =\const$, $s \geq 0$.

  The definitive property of vacuum matter is that the SET (\ref{SET-v})
  preserves its form under arbitrary radial boosts \cite{dym}, hence,
  unlike ordinary matter, it has no distinguished comoving reference frame.
  Examples of such matter are usual Maxwell radial electric and magnetic
  fields for which $p_\bot\vac = \rho\vac$, their analogs in nonlinear
  electrodynamics with Lagrangians of the form $L_e = L_e(F)$, $F\equiv F\mn
  F\MN$ \cite{br-NED}, Yang-Mills fields with a similar structure of the
  SET, and clouds of radially directed cosmic strings \cite{letelier79}.

  For matter that we here call pseudo-vacuum, the condition $\rho + p_r =0$
  holds only at the horizon, and its properties are quite arbitrary in
  all the remaining space.

  As already mentioned, a true vacuum fluid is compatible with horizons of
  any order. Pseudo-vacuum matter is much more general, as is seen, in
  particular, from (\ref{ps1}) and (\ref{rho_e1}), therefore it is clear
  that any horizons can be in equilibrium with it, and a detailed analysis
  along the same lines as before reveals more than a dozen of
  particular forms of such an equilibrium. We, however, mention here
  just two features that can be of certain interest:

\begin{itemize}
\item
     Anisotropy of pseudo-vacuum matter is required to provide equilibrium
     with horizons of orders $n > 2$.
\item
     In all cases where the matter charge density is significant for
     maintaining the equilibrium, we have $s > 0$, i.e., the density
     $\rho_e$ vanishes at the horizon (though slowly enough).
\end{itemize}

%%%%%%%%%%%%%
\subsection {A non-interacting mixture of ordinary and (pseudo)vacuum
	matter}.

  Let us now find out what changes in our conclusions made in Sections
  3.1--3.3 if we add to the system described there a vacuum or pseudo-vacuum
  fluid with the SET
\beq                                                     \label{SET-v}
      T\mN\vac = \diag (\rho\vac, \rho\vac, -p_\bot\vac, -p_\bot\vac),
\eeq
  and there is no interaction between the SETs $T\mN{}^m + T\mN{}^{\rm em}$
  and $T\mN\vac$, that is, the conservation law (\ref{cons}) holds, and
  we also have $\nabla_\nu T\mN\vac =0$.\footnote
  	{In our reasoning here, a difference between vacuum and
	pseudo-vacuum matter is insignificant because only the values of its
	density and pressures at the horizon will be relevant.}
  If the SET (\ref{SET-v}) is added to (\ref{T_m}) and (\ref{T_em}), the
  conservation law (\ref{cons}) for matter under consideration and the
  Einstein equation (\ref{01}) do not change (and (\ref{01}) again leads to
  $k \geq n$), while (\ref{11}) and (\ref{02}) take the form
\bear                                           		 \label{11v}
      G_1^1 \eql \frac 1 {r^2}[-1 + A'rr' + A r'{}^2]
			 = 8\pi p_r - \frac{e^2}{r^4} -8\pi \rho\vac,
\yy                                                             \label{02v}
      G^0_0 -G^2_2 \eql \frac{1}{2r^2}[-2 +A (r^2)'' - A''r^2]
		  	= - 8\pi (\rho + p_\bot) - \frac{2e^2}{r^4}
		  	  - 8\pi (\rho\vac+ p_\bot\vac).
\ear

  \eq (\ref{11v}) in the order $O(1)$ gives
\beq                                                            \label{11v0}
       1 - A'rr'\Big|_{u=0} = \ebh^2/r_h^2 + \Lambda r_h^2,
  		\cm \Lambda := 8\pi\rho\vac\Big|_{u=0}.
\eeq
  Therefore higher-order horizons ($n > 1$), such that $A'(0)=0$, become
  possible even for $\ebh = 0$.

  \eq (\ref{02v}) in the same order of magnitude leads to
\beq                                                           \label{02v0}
       2 + A''r^2\Big|_{u=0} = 4\ebh^2/r_h^2 + 2\Lambda_* r_h^2,
       \cm
  	\Lambda_* ;= 8\pi(\rho\vac + p_\bot\vac)\Big|_{u=0}.
\eeq
  Now it is easy to obtain equilibrium conditions in the presence of
  vacuum matter.

\medskip\noi
{\bf (a)} The equation of state (\ref{w}), $\ebh=0$ (a neutral \bh). Horizons
  of any order $n$ are possible, and (\ref{cons}) gives
\beq                                                         \label{w_n}
	w = - \frac{n}{2k + n},
\eeq
  reproducing the result of \cite{curly}. \eqs (\ref{11v0}) and (\ref{02v0})
  lead to relations between the parameters of the configuration. For $n > 1$
  we obtain
\beq
	\Lambda r_h^2 = 1, \cm (2\Lambda_* - A''(0))r_h^2 = 2,
\eeq
  and the second relation indicates that horizons of order $n > 2$ are only
  possible if $\Lambda_* > 0$. The cosmological constant, for which
  $p_\bot\vac = p_r\vac = -\rho\vac$, is thus excluded.

  A possible nonzero $\rho_e$ is insignificant.

\medskip\noi
{\bf (b)} The equation of state (\ref{w}), $\ebh \ne 0$. Again horizons of
  any order are possible; \eq (\ref{cons1b}) is again valid and leads to
\beq
	s \geq n/2 -1.                                      \label{s>_}
\eeq
  In the case of a simple horizon, as before, the last term in (\ref{cons})
  is insignificant, so that for $w$ we have the old result (\ref{w1a}), and
  it is required that $s$ should be half-integer.

  For $n \geq 2$ and the exact equality in (\ref{s>_}) (in particular,
  $s = 0$ for $n = 2$), we obtain instead of (\ref{w1b})
  the relation
\beq                                                       \label{w1v}
	w = \frac{2\alpha_* - n}{2k + n},
	\cm \alpha_* := \frac{\alpha \ebh}{r_h^2 \sqrt{A_0}},
\eeq
  while due to (\ref{11v0}) and (\ref{02v0}) the parameters of the
  configuration are related by
\beq
	\ebh^2/r_h^2 = 1 - \Lambda r_h^2,
    \cm                                                    \label{par1v}
	\delta_{n2} A_0 r_h^2 = 1 + r_h^2 (\Lambda_* - 2\Lambda)
\eeq
  (the Kronecker symbol $\delta_{n2}$ indicates that this term appears only
  if $n=2$).

  If (\ref{s>_}) is a strict inequality (i.e., our matter is weakly
  charged), we must put $\alpha_*=0$ in (\ref{w1v}) and thus return to the
  old expression (\ref{w_n}).

\medskip\noi
{\bf (c)} Neutral matter with the equation of state (\ref{m}). The
  conclusion made in Sec. III.B, that such matter cannot be in equilibrium
  with any \bh, rests solely on \eq (\ref{cons}) and therefore does not
  change in the presence of a vacuum fluid.

\medskip\noi
{\bf (d)} Charged matter with the equation of state (\ref{m}). As in
  Sec.\,III.C, the conservation law near the horizon leads to (\ref{cons3})
  which again discards simple horizons and yields $s = n/2 - 1$ for
  higher-order horizons ($n \geq 2$).

  Unlike Sec.\,III.C, horizons of any order $n\geq 2$ now become possible, 
  the system parameters obeying the relations (\ref{par1v}), and instead of
  (\ref{al-c}) we now find the condition
\beq
       \alpha_*  = n/2.
\eeq
  The power $m$ in (\ref{m}) remains arbitrary but must obey the condition
  $mk \in \N$ to maintain analyticity.

\section {Conclusion}
\def\Vac {\mbox{\sf vac }}

  Our main results are summarized in the following table.

\newpage  \noi      
       {\small {\bf Table:}
	Matter equilibrium conditions for neutral and charged black holes
	with horizons of order $n$. The symbol ``{\Vac}'' denotes 
	(pseudo)vacuum matter, the symbol $\oplus$ a non-interacting mixture, 
	and $\{s\}$ is the fractional part of $s$. 
	All other notations can be found in the text.}

\bigskip\noi
\begin{tabular}{|c|l|l|} 
\hline 
	Matter  & \  $\ebh = 0$     & \ $\ebh \ne 0$ \\
\hline \hline
	\Vac    & \  $\forall \ n$  & \ $\forall \ n$ \\
\hline
     \wide &&  \        $n=1, \ \ w = -\dfrac 1{1+2k}, \ \ \{s\}=1/2$  \\
     \wide $p_r = w\rho$ & \ $n=1, \ \ w = -\dfrac 1{1+2k}$\ & \
		    $n=2, \ \ w =  \dfrac {\alpha'-1}{1+2k}, \ \ s=0$ \\
     \wide &&  \       $n=2, \ \ w = -\dfrac 1{1+k}, \ \ \{s\}=1/2$    \\
\hline
     \wide $ \Vac \oplus (p_r = w\rho)$  & \
		     $\forall\ n, \ \ w = -\dfrac n{2k+n}$  & \
			$n=1, \ \ w = -\dfrac 1{1+2k}, \ \ \{s\}=1/2$  \\
     \wide && \ $n\geq 2, \ \ w = -\dfrac {2\alpha_*-n}{1+2k}, \ \ s=n/2-1$\\
\hline
     $p_r = B\rho^m, \ m>1$ & \    none    & \
		    $n=2, \ \ \alpha = \pm 1,\ \ s=0$    \\ 				
\hline
     $\ \Vac \oplus (p_r = B\rho^m, \ m>1)$\	& \ none  & \
		    $n \geq 2,\ \ \alpha_* =n/2, \ \ s=n/2-1$ \\  
\hline
\end{tabular} 

\bigskip\bigskip
  Thus, in the framework of \ssph\ configurations in general relativity, we 
  have analyzed the equilibrium conditions between an electrically charged
  or neutral \bh\ and ambient matter. To do so, we used near-horizon
  expansions of the Einstein equations and the conservation law for matter.
  Actually, all the conditions were obtained from the largest terms in such
  expansions; in general, smaller terms do not lead to new
  restrictions but only provide relations between further expansion factors
  of the corresponding functions of the radial coordinate.

  For linear equations of state of the form $p_r =w\rho$, we have 
  confirmed our previous results \cite{curly} that in the case of neutral 
  matter an equilibrium is possible for a discrete set of values of $w$ 
  between zero and -1/3 ($w = -1/3$ is the generic case, corresponding to 
  a distribution of disordered cosmic strings if the matter is isotropic).
  We have found out how these conditions are modified in the presence of 
  an electric charge. It was stressed in \cite{curly} that inside a 
  ``normal'' star (i.e., with everywhere nonnegative pressure) a black hole 
  cannot exist. However, as follows from the present results, overcharged 
  normal matter ($w >0$, $\alpha > 1$) can be in equilibrium with a \bh\ 
  but its horizon must be extremal ($n = 2$).

  Neutral matter with a nonlinear equation of state (\ref{m}) cannot be in
  equilibrium with any \bh. However, charged matter with the same equation
  of state can surround extremal charged \bhs\ under the same condition
  on the charge density as is known for the Majumdar-Papapetrou solutions.

  Some additional possibilities of this kind are opened if vacuum 
  (or pseudo-vacuum) matter is added to the system, and in particular, 
  only in its presence horizons of orders $n >2$ (ultraextremal) become 
  possible.

  Thus inclusion of an electric charge (however small it be) into
  consideration leads to qualitatively new possibilities for both linear 
  and nonlinear equations of state which were absent in the neutral case.

  Our reasoning has been entirely local and relied on near-horizon 
  expansions, thus no assumptions on global or asymptotic properties 
  of space-time have been made. We also did not assume any particular 
  equation of state for matter and even did not restrict 
  the behavior of the transverse pressure except for its regularity 
  requirement. In this sense, our conclusions are model-independent.

\Acknow
   {The authors thank Jos\'e P.S. Lemos for a helpful discussion.
    The work of K.B. was supported in part by the Russian Foundation for
    Basic Research Grant No. 09-02-00677a, by NPK MU grant at PFUR, and
    by FTsP ``Nauchnye i nauchno-pedagogicheskie kadry innovatsionnoy
    Rossii'' for the years 2009-2013.}

\end{document}